\documentclass[aps,prb,superscriptaddress,floatfix,preprint]{revtex4-1}
\usepackage{amssymb}
\usepackage{amsmath}
\usepackage{graphicx}
\usepackage{color}

\begin{document}

\title{\Huge Separation of neutral and charge modes in one dimensional chiral edge channels}

\author
{E. Bocquillon,$^{1}$ V. Freulon,$^{1}$ J.-M Berroir,$^{1}$  P. Degiovanni,$^{2}$ \\
B. Pla\c{c}ais,$^{1}$ A. Cavanna,$^{3}$ Y. Jin,$^{3}$ G. F{\`e}ve$^{1\ast}$ \\
\normalsize{$^{1}$Laboratoire Pierre Aigrain, Ecole Normale Sup\'erieure, CNRS (UMR8551), Universit\'e Pierre et Marie Curie, Universit\'e Paris Diderot}\\
\normalsize{24 rue Lhomond, 75231 Paris Cedex
05, France}\\
\normalsize{$^{2}$ Universit\'e de Lyon, F\'ed\'eration de Physique Andr\'e Marie Amp\`ere,} \\
\normalsize{CNRS - Laboratoire de Physique de l'Ecole Normale Sup\'erieure de Lyon} \\
\normalsize{46 All\'ee d'Italie, 69364 Lyon Cedex 07,France.}\\
\normalsize{$^{3}$CNRS - Laboratoire de Photonique et de Nanostructures}\\
\normalsize{Route de Nozay, 91460 Marcoussis, France}\\
\normalsize{$^\ast$ To whom correspondence should be addressed;
E-mail:  feve@lpa.ens.fr.} }

\begin{abstract}
Coulomb interactions have a major role in one-dimensional electronic transport. They modify
the nature of the elementary excitations from Landau quasiparticles in higher dimensions to
collective excitations in one dimension. Here we report the direct observation of the collective
neutral and charge modes of the two chiral co-propagating edge channels of opposite spins of
the quantum Hall effect at filling factor 2. Generating a charge density wave at frequency $f$ in
the outer channel, we measure the current induced by inter-channel Coulomb interaction in
the inner channel after a 3-mm propagation length. Varying the driving frequency from $0.7$ to
$11$ GHz, we observe damped oscillations in the induced current that result from the phase shift
between the fast charge and slow neutral eigenmodes. We measure the dispersion relation
and dissipation of the neutral mode from which we deduce quantitative information on the
interaction range and parameters.
\end{abstract}
\date{\today}
\maketitle

\newcommand{\CC}{\mathcal{C}}

\maketitle

Most studies of collective excitations in one dimensional systems
performed so far have focused on non-chiral quantum wires
\cite{Auslaender2005, Steinberg2007,Jompol2009}. In these systems,
the collective excitations carrying the charge and the spin
propagate at different velocities leading to the separation of the charge and spin degrees of freedom.
This spin-charge separation has been probed by measuring the tunneling spectroscopy of individual
electrons between a pair of one dimensional wires \cite{Auslaender2005}, or
alternatively, between a wire and a two dimensional electron gas \cite{Jompol2009}.
However, a direct observation of the collective modes is experimentally challenging as the relevant energy scales are too high
for usual low frequency measurements.

The edge channels of the quantum Hall effect provide another implementation of one
dimensional transport, where propagation is chiral and ballistic
over large distances. These specificities have inspired several experiments that aim at reproducing, in solid state,
optical setups where light beams are replaced by electron
beams\cite{Henny1999,Ji2003,Bocquillon2012,Bocquillon2013}. One
major difference between electrons and photons comes from
interaction effects, which are amplified in the one dimensional
geometry and should enrich electron optics compared to its
photonic counterpart. Of particular interest is the case of filling factor $2$
where transport along the sample edge occurs through two
copropagating edge channels of opposite spins. Due to Coulomb
interaction, the two edge states are coupled and new propagating
eigenmodes, with different velocities,
appear\cite{Lee1997,Sukhorukov2007,Levkivskyi2008,Berg2009,Levkivskyi2012}, similarly to the physics arising in 1D wires. Consequently,
considering edge channels with the same propagation
characteristics (but different spins), and denoting
$\vec{i}(x,t)=(i_1(x,t),i_2(x,t))$ the current components in edge channels $1$
and $2$ at position $x$ and time $t$,  a current $(i_1,0)$ in
channel $1$ decomposes in the symmetric fast charge mode
$(\frac{i_1}{2},\frac{i_1}{2})$ and the antisymmetric
slow neutral mode $(\frac{i_1}{2},-\frac{i_1}{2})$
(also called dipolar mode). As these two modes propagate at
different velocities, the current initially injected in channel 1 separates into the charge and
the neutral modes of the two coupled edges . This mechanism is at the heart of the
decoherence\cite{Sukhorukov2007,Levkivskyi2008,Grenier2011} and
relaxation\cite{Kovrizhin2010,Degiovanni2010,Lunde2010,Levkivskyi2012} of electronic
excitations propagating in these systems. As such, it has been
probed through Mach-Zehnder
interferometry\cite{Neder2006, Roulleau2008,Huynh} or spectroscopy of edge
channels \cite{Altimiras2009,Altimiras2010,LeSueur2010}. This situation bears strong analogies with the spin-charge
separation of conventional 1D wires except that the two spin species are carried by two separated edge channels.
The direct observation of
the neutral and charge eigenmodes is thus particularly favorable in the filling factor 2 case as, contrary to a conventional wire, each spin
channel can be individually addressed due to their spatial
separation. Still, the observation remains challenging as transport properties
remain unaffected up to GHz frequencies,
where the wavelength of the eigenmodes becomes comparable with the
propagation length of a few microns. Many experimental works
have studied charge transport in quantum Hall edge channels and interaction effects between copropagating and counterpropagating edge channels either in time \cite{Ashoori1992,Zhitenev1993,Ernst1996,Sukhodub2004} or in frequency \cite{Gabellia,Talyanskii1992,Hashisaka2012,Andreev2012} domains. However, none directly addressed the separation in charge
and neutral modes as the individual control of edge channels was missing.

In this work, by addressing edge channels individually, we provide a direct observation of the neutral-charge eigenmodes. Using a driven
mesoscopic capacitor (Fig.\ref{fig1} (a), see Methods for
details), a sinusoidal charge density wave, or edge magnetoplasmon
(EMP) is induced  at pulsation $\omega$ and position $x=0$ in
channel 1, thus creating a current $i_1(x=0, \omega)$ (with
$i_1(x,t)=i_1(x,\omega)e^{-i\omega t}$). As initially introduced in the context of a non chiral one dimensional wire  \cite{Safi1995} and later developed for chiral edge channels \cite{Degiovanni2010} at filling factor $2$, Coulomb interaction during propagation can be described as the scattering of the charge density waves. For the case of filling factor $2$ of interest here, scattering properties of EMPs are encoded in the $2\times2$ scattering matrix ${\bf S}$ that relates the amplitudes of the output EMP in channels 1 and 2 after propagation length $l$ to the
amplitudes of the input EMP at $x=0$. After an interaction length $l=3.2\pm0.4\, \mu$m, both edge channels reach a
quantum point contact (QPC) which is used to transmit or reflect channels 1 and 2.
Figure \ref{fig1}(b) presents the principle of measurement for two
typical sets of data (for $f=1.3$ and 5.5 GHz). In configuration $1$, channel 1
is transmitted and channel $2$ is reflected.
The current in channel 2 resulting from the
interaction, denoted $i_2(l,\omega)$, can then be measured in ohmic contact $A$, with
$i_2(l,\omega)=S_{21}(l, \omega)\,i_1(0,\omega)$. When the QPC is
closed (configuration 2), both channels are reflected so that the
total collected current in $A$ is $
i_1(l,\omega)+i_2(l,\omega)=
\big(S_{11}(l,\omega)+S_{21}(l,\omega)\big)\,i_1(0,\omega) $.
Consequently, the ratio of the currents collected in these two
configurations yields the complex quantity
$\mathcal{R}(\omega) =\frac{S_{21}(\omega)}{S_{11}(\omega)+S_{21}(\omega)}$, which encodes the
effect of Coulomb interaction on the propagation along the edge
states.

\section*{Results}

\subsection*{Inter-edge oscillations of EMP}

The experimental data for $\mathcal{R}$ are presented in figure
\ref{fig2} (colored dots), in the complex plane. The color code
gives an insight of the driving frequency $f$. Globally, we
observe that $\mathcal{R}$ draws a spiral in the complex plane.
For $f=0.7$ GHz, we observe that $\mathcal{R}\simeq 0$, reflecting
the fact that the current injected in the outer channel remains in
this channel ($S_{21}(\omega=0)\simeq 0$). At low frequencies
($f<2$ GHz) , $\mathcal{R}$ is mainly imaginary, as expected for a
capacitive coupling between the two edge states. As frequency increases,
$\mathcal{R}$ winds around, reaching a maximum $|\mathcal{R}|\simeq0.75$ for
$f\simeq4.5$ GHz, meaning that 75\% of the charge density wave has
been transferred to the inner channel. For increasing $f$,
$\mathcal{R}$ continues to spiral, decreasing to
$|\mathcal{R}|\simeq0.4$, then increasing again above 9 GHz (for
clarity, the corresponding data are not shown on Fig.2 but the modulus and phase of $\mathcal{R}$ in the range 9 GHz $\leq f \leq$ 11 GHz are shown on Fig.\ref{fig3bis}). This behavior
demonstrates coherent oscillations of the charge density wave from one edge to the
other as the driving frequency $f$ is varied. These
oscillations can be understood in a simple manner. Taking here as
eigenmodes (this assumption is discussed later) the symmetric charge mode
$\vec{u}_\rho=(\frac{1}{\sqrt{2}},\frac{1}{\sqrt{2}})$ and the
antisymmetric neutral mode $\vec{u}_n=(\frac{1}{\sqrt{2}},-\frac{1}{\sqrt{2}})$, the current $\vec{i}$ can be decomposed in the eigenbasis as $
\vec{i}=i_\rho\vec{u}_\rho+i_n\vec{u}_n$, where $i_{\rho/n}$ are the coordinates in the eigenbasis, $i_{\rho/n}=\frac{1}{\sqrt{2}}(i_1\pm i_2)$.
Their propagation reduces to a phase factor $S_{\rho/n}=e^{i\frac{\omega
l}{v_{\rho/n}}}$, where $v_{\rho/n}$ are the phase velocities of respectively the charge and neutral modes, with:
\begin{eqnarray}
i_{\rho/n}(l,\omega)&=&S_{\rho/n}(l,\omega)\, i_{\rho/n}(0,\omega)\\
S_\rho&=&S_{11}+S_{21}=e^{i\frac{\omega l}{v_\rho}}\label{Srho}\\
S_n&=&S_{11}-S_{21}=e^{i\frac{\omega l}{v_n}}\label{Sn}\\
\mathcal{R}&=&\frac{1-e^{i\omega l(\frac{1}{v_n}-\frac{1}{v_\rho})}}{2}\simeq\frac{1-e^{i\frac{\omega l}{v_n}}}{2}\simeq S_{21}\label{S21velocity}
\end{eqnarray}
where we have assumed in  Eq.(\ref{S21velocity}) that the charge mode propagates much faster than the neutral mode,
$v_n \ll v_\rho$, such that $\frac{\omega l}{v_\rho}\ll 2\pi$. The term $e^{i\omega
l(\frac{1}{v_n}-\frac{1}{v_\rho})}$ in Eq.(\ref{S21velocity})
shows that the oscillation stems from the progressive phase
shift between the charge and neutral components of the EMP
propagating at different velocities. In the
complex plane, $\mathcal{R}$ should describe a circle of radius 1/2 centered
at $(1/2, 0)$, with angle $\Phi(\omega)=\frac{\omega
l}{v_n(\omega)}-\pi$ when the frequency is varied. At low frequency, the propagation length $l$ is much smaller than the wavelength
of the neutral mode $\lambda_n=2 \pi v_n/\omega$ and propagation effects can be neglected. One then recovers the well-established\cite{Gabelli2006a,Roulleau2008a} RC-circuit
limit where $\mathcal{R}$ starts from $\mathcal{R}(\omega=0)=0$ and $\mathcal{R}(\omega)=-i\omega\tilde\tau(1+i\omega\tilde\tau)$ for
$\omega \tilde\tau \ll 1$, with
$\tilde\tau=\frac{l}{2 v_n}$. $\tilde\tau$ can be expressed in term of discrete elements $R_K$ and $C_{\mu}$, $\tilde\tau=R_K C_{\mu}$ (see Methods).
$C_{\mu}$ is the electrochemical capacitance given by the series
association of a quantum capacitance
$C_q=\frac{l}{R_K v}$ for each channel (where $v$
is the velocity in the edge channels in the absence of inter or intra edge channel interactions) and the geometrical
capacitance between channels $C$. The resistor is $R_K=h/e^2$, the series combination of a
charge relaxation resistance $R_K/2$ for each channel. This
corresponds to the low frequency velocity
$v_n^0=v+\frac{l}{2R_KC}=v(1+\frac{C_q}{2C})$. At higher frequencies, the propagation length becomes comparable with the wavelength
of the neutral mode and propagation effects cannot be ignored. Using Eq.(\ref{S21velocity}),
the trajectory followed by $\mathcal{R}$ in the complex plane then gives a direct access to the neutral mode velocity $v_n(\omega)$ or, in an equivalent way, to the $\omega$ dependence of the wave vector $k_n(\omega)$ (the dispersion relation) related to the phase velocity by
$k_n(\omega)=\frac{\omega}{v_n(\omega)}$. One can see on Fig.\ref{fig2} that when the frequency is increased, $\mathcal{R}$ follows the expected
circle for $f<4$ GHz (Eq.(\ref{S21velocity}) is plotted in black line).  However, for frequency ranging from 4 to 9 GHz, data points deviate from the expected circle and the
experimental curve seems to spiral down towards a state where the
charge density wave is evenly distributed between both channels with
$\mathcal{R}\to \frac{1}{2}, S_{11}\simeq S_{21}\to\frac{1}{2}$.
This can be understood as a dissipation of the EMP during propagation and can be accounted for by introducing an
imaginary part in the wave vector $k_n(\omega)$. These considerations bring to light the remarkable robustness of
the Nyquist diagram presented on Fig.\ref{fig2}. Separation between the charge and neutral mode show up in the inter-edge oscillations
revealed by the winding of $\mathcal{R}$
around the point $(1/2,0)$, which corresponds to an equal repartition of the EMP.
This feature, clearly visible on Fig.\ref{fig2} in spite of dissipation, does not depend on the
details of the interaction.  The interaction characteristics are  encoded in the dispersion relation
$k_n(\omega)$ or equivalently in the $\omega$-dependence of
the phase $\Phi(\omega)$. To depict interactions, the most
frequently used approach is a zero-range
model\cite{Talyanskii1992,Sukhorukov2007,Berg2009}. It has no
characteristic length, so that the velocity $v_n$ is
frequency independent:
$v_n(\omega)=v_n^0=v(1+\frac{C_q}{2C})$. Since
$v_n$ is frequency independent, $S_{21}$ draws a circle with
a linear $\omega$-dependence of the phase $\Phi(\omega)\propto
\omega$. Any deviation from this linear dependence reflects the
existence of a finite range in the interactions which can be unveiled through a careful study of
the frequency dependence of $\mathcal{R}$.

Fig.\ref{fig3} presents our measurements of the dispersion relation $k_n(\omega)$ where
the complex value of $k_n$ is extracted from  Eq.(\ref{S21velocity}). As already mentioned, the existence of an imaginary part ${\rm Im}(k_n)$ in the
wave vector $k_n$ signals the presence of dissipation. Two
non-dispersive regimes are observed : at low frequency, ${\rm
Re}(k_n)=\frac{\omega}{v_n(0)}$, with
$v_n(0)=4.6 \pm 0.6 \times 10^4$ m.s$^{-1}$. This regime of constant velocity with respect to frequency is
consistent with a short range description of interactions in the low frequency limit. However, for $f>6$ GHz, a
second linear dispersion relation regime appears ${\rm
Re}(k_n)=\frac{\omega}{v_n(\infty)}$, with
$v_n(\infty)=2.3 \pm 0.3 \times 10^4$ m.s$^{-1}$. We attribute this
decrease of $v_n$ to the finite range of interactions. To go beyond this qualitative discussion, we now rely on a quantitative comparison
between our experimental data for $\mathcal{R}$  and various models of intra-edge and inter-edge interaction.

\subsection*{Comparison between model and data}

In figure \ref{fig3bis}, experimental data for $|\mathcal{R}|$ and $\arg(\mathcal{R})$
are presented  as a function of driving frequency $f$. At low
frequency, the RC-circuit behavior is recovered and the agreement
with the experimental data is good in the range 0.7--3 GHz, with a
value of  $\tilde\tau=\frac{l}{2v_n(0)}=35$ ps extracted from the low-frequency regime of the dispersion relation. However, this
RC-circuit model (green small dashes) does not predict
oscillations. The zero-range interaction
model, obtained by locally coupling the electrostatic potential of the edges to the charge densities
(see Methods and Fig.\ref{fig5}) for which the velocity is frequency independent, $v_n=v_n^0=l/(2 \tilde\tau)$ is plotted in dashed red line.
As already discussed, it does predict the oscillations, but fails
to describe accurately the regime of high frequencies (3 to 11
GHz). Once $\tilde\tau$, which prescribes the low
frequency behavior, is fixed, no other parameter is to be fixed for
the zero range model.

To capture these features, an heuristic long range model has been
developed (see Methods and Fig.\ref{fig5}) based on a discrete element description
\cite{Buttiker93,Christen1995}. An effective range on the order of the
propagation length $l$ is introduced by assuming that the
electrostatic potentials of each edge state are constant over $l$,
and by coupling these potentials to the total charges in the
channels via a capacitance $C$. This long range model depends on
two timescales $\tau = R_K C$ and $\tau_q=R_KC_q=l/v$ such that the
velocity  $v_{n}(\omega)$ becomes frequency dependent,
$v_n(\omega \to 0)=\frac{l}{2\tilde\tau}=v_n^0$ (with
$\tilde\tau =\frac{\tau\tau_q}{2\tau+\tau_q}$) and
$v_n(\omega \to \infty)=\frac{l}{\tau_q}=v$. An intrinsic dissipation
inside each edge channel that accounts for the damping of the charge oscillations
 is introduced through the parameter
$\gamma(\omega)$. We choose the following $\omega$ dependence, $\gamma(\omega) = \omega^2 \tau_r$,
which is still compatible
with a discrete circuit elements description at low frequency. Dissipation modifies the value of the resistance in the RC circuit description :
a resistor of $R_r=\frac{R_K}{2}\frac{\tau_r}{\tilde\tau}$ is added in series
with the charge relaxation resistor $R_K/2$. Putting first $\tau_r=0$ (no dissipation), this
long-range interaction model captures already both the
low-frequency behavior and the period of oscillation (see the blue
dash-dotted curve on Fig.\ref{fig3bis}) from which we extract
$\tau_q= 124\pm5$ ps (which implies $\tau=81\pm4$ ps since we have set $\tilde\tau=35$ ps). A good
agreement with experimental data including dissipation can then be
obtained using $\tau_r=4.1\pm1$ ps (black curve on Fig. \ref{fig3bis}). Note that the value of $\tau_r$
extracted  from our data is rather small, $\tau_r \simeq\frac{\tilde\tau}{9}$ such that the
low-frequency regime is not strongly affected. From the evaluated value of $\tau_q$, we also deduce $v=\frac{l}{\tau_q}=2.6 \pm 0.2 \times 10^4$
m.s$^{-1}$, consistently with the velocity $v_n(\infty)=2.3 \times 10^4$
m.s$^{-1}$ extracted from the
dispersion relation. These values are compatible with the assumption $v_n\ll v_\rho$ estimating the charge velocity
from experiments performed with similar samples\cite{Kumada2011}. In ref. \onlinecite{Kumada2011}, Kumada {\it et al.} indeed find a charge velocity of a few $10^6 $ m.s$^{-1}$ at filling factor $2$ for an ungated two dimensional electron gas. Their sample characteristics are close to ours, it is made from a Gallium Arsenide heterostructure,
the electron gas has a density of $1.2 \times 10^{15}\,\rm m^{-2}$ and mobility $\mu= 2.1 \times 10^6\, \rm cm^2 V^{-1}s^{-1}$
(close to our values, see Methods) and the sample edges are defined by chemical etching (as ours).
Simulations of the dispersion relation with the same parameters are also presented in Fig.\ref{fig3}. The overall behavior of Re($k_n$) is well-rendered:
though not as abrupt, the change in the velocities is as expected described by the long-range interaction.
In the meantime, Im($k_n$) is also correctly depicted with our choice of dissipation for the EMP: $\gamma(\omega)=\omega^2\tau_r$.

\subsection*{Nature of the eigenmodes}

Throughout the paper, the case of symmetric edge channels has been considered which naturally leads to the existence
of pure charged and neutral eigenmodes. As there is no reason for both edge channels to have identical propagation properties, one should consider
in full generality the decomposition $\vec{i}=i_+\vec{u}_+ + i_-\vec{u}_-$, where the eigenmodes $\vec{u}_+$ and $\vec{u}_-$ are parametrized by the angle $\theta$ :
 \begin{eqnarray}
 \vec{u}_+&=&\left( \cos\frac{\theta}{2},\sin\frac{\theta}{2} \right),\quad \vec{u}_-=\left(\sin\frac{\theta}{2},-\cos\frac{\theta}{2}\right)\label{EigenVecs}\\
 i_+(x,\omega) & = & \cos{\frac{\theta}{2}} \; i_1(x,\omega) + \sin{\frac{\theta}{2}} \; i_2(x,\omega) \label{EigenPlus} \\
 i_-(x,\omega) & = & \sin{\frac{\theta}{2}} \; i_1(x,\omega) - \cos{\frac{\theta}{2}} \; i_2(x,\omega)\label{EigenMinus} \\
 i_{\pm}(l,\omega) & = & e^{i \frac{\omega l}{v_{\pm}} } \; i_{\pm}(0,\omega)\label{EigenPhase}
 \end{eqnarray}
 The case $\theta=0$ corresponds to completely independent channels while $\theta= \pi/2$ corresponds to the strong coupling
 case where the eigenmodes are the charged ($+=\rho$) and neutral ($-=n$) modes. As already discussed, the latter situation occurs in the case of identical edge
 channels but can also occur for non-identical channels as long as the interchannel interaction is strong enough (see Methods).
 Any other intermediate case corresponds to partially charged eigenmodes for which one can define the ratio $r$ of the total charge carried by modes $-$ and $+$ from the expression of the eigenmodes, Eq.(\ref{EigenVecs}) : $r=\frac{\sin{\frac{\theta}{2}} - \cos{\frac{\theta}{2}}}{\cos{\frac{\theta}{2}} +\sin{\frac{\theta}{2}} }$. For $\theta=\pi/2$, the contribution of the antisymmetric mode to the current is 0, reflecting its neutrality.
 In this general case, the expressions for $S_{21}$ and $S_{11} + S_{21}$, and thus for the measured quantity $\mathcal{R}$, differ from Eqs. (\ref{Srho}), (\ref{Sn}), (\ref{S21velocity}):
 \begin{eqnarray}
 S_{21} & =& \sin{\theta } \; \frac{e^{i\omega l/v_+} - e^{i\omega l/v_-}}{2} \label{S21Theta}\\
 S_{11} + S_{21} & =& \frac{e^{i\omega l/v_+} + e^{i\omega l/v_-}}{2} + (\cos{\theta} +  \sin{\theta })  \frac{e^{i\omega l/v_+} - e^{i\omega l/v_-}}{2} \label{SsumTheta}\\
 \mathcal{R} & = &  \frac{\sin{\theta } \; (1 - e^{i \phi(\omega)})} {1  + e^{i \phi(\omega)} + (\cos{\theta} +  \sin{\theta })\; (1 - e^{i \phi(\omega)} )} \label{RTheta}\\
 \phi(\omega) & = & \frac{\omega l}{v_-} -\frac{\omega l}{v_+}
 \end{eqnarray}
 Eq.(\ref{S21Theta}), shows that $S_{21}$ describes coherent oscillations from channel 1 to channel 2, whose amplitude are
given by the factor $\sin{\theta}$. This amplitudes only reaches unity at strong coupling $\theta=\pi/2$, the only
regime where a complete charge transfer from one channel to the other can be achieved.
$S_{11}+S_{21}$ is also affected and oscillates (either with frequency $f$ or length $l$)
 reflecting the fact that in this general case, the charge mode $\vec{u}_{\rho}$ is no longer an eigenmode such that the total current $i_1 + i_2$ oscillates instead of simply accumulating a phase. As a result, $\mathcal{R}$ still follows a
 circle in the complex plane but with a $\theta$ dependent radius and center.
 From these general expressions and the comparison with our experimental data, one can assess that the eigenmodes are indeed the charge and neutral ones, within an accuracy of $r=0 \pm 0.1$ for the charge ratio between the eigenmodes. The first argument comes from the low frequency
 behavior of $\mathcal{R}$ where dissipation can be safely neglected. Both the modulus $|\mathcal{R}|$ and the phase $\arg(\mathcal{R})$ follow a linear $\omega$
 dependence but with two different $\theta$ dependent slopes, $|\mathcal{R}| = \sin{\theta}\; \phi(\omega)/2$, $\arg(\mathcal{R}) = -\frac{\pi}{2} +  \big(\sin{\theta} + \cos{\theta} \big) \; \phi(\omega)/2$.
 By measuring the ratio of these slopes, one can directly measure the angle $\theta$. Remarkably, in the strong coupling case, $\theta = \frac{\pi}{2}$, data points for $|\mathcal{R}|$ and $\arg(\mathcal{R})+ \frac{\pi}{2}$
 should follow the exact same frequency dependence in the low frequency regime. Data points in the low frequency $0.9 - 4.5$ GHz range are plotted on Fig.\ref{fig6}, with their linear fits. A linear fit for $|\mathcal{R}|$ (in the $0.9 - 2$ GHz range, as $|\mathcal{R}|$ approaches its maximum for $f\gtrsim 3 $ GHz) is presented in black plain line, yielding $|\mathcal{R}| = 3.8 \times 10^{-11} \times \omega$.
Similarly, $\arg(\mathcal{R})+ \frac{\pi}{2}$ is fitted in red line in the range $1.2 - 4.5$ GHz range, with $\arg(\mathcal{R})+ \frac{\pi}{2} = 3.7 \times 10^{-11} \times \omega$. Data below 1.2 GHz are not used in the fitting procedure due to their dispersion, as, at low frequency, $\arg(\mathcal{R})$ is obtained from the ratio between two small currents.
 The slope of $|\mathcal{R}|$, that is, the value of $\sin\theta \times\frac{\phi(\omega)}{\omega}$ for small $\omega$, is determined with a 10\% accuracy. It thus defines two bounds (dashed blue lines) that correspond to the slopes of $\arg(\mathcal{R})+ \frac{\pi}{2}$ for $\theta=90 + 11^\circ$ (upper bound)  $\theta=90 - 11^\circ$ (lower bound). These extremum values of the angle $\theta$ correspond to a charge ratio  $r = \pm0.1$.  Our data points for  $\arg(\mathcal{R})+ \frac{\pi}{2}$ fall between these bounds which assesses the neutrality of the slow mode with a 10\% accuracy.
 The second argument comes from the study of the full trajectory of $\mathcal{R}$  in the complex plane. From the amplitude of the oscillations of the EMP from one edge to the other, a lower bound for $\theta$ can be obtained , $\theta \geq 74^\circ$ corresponding to $|r| \leq 0.15$. This lower bound is obtained by assuming that the amplitude of the oscillation is only limited by the value of $\theta$ and neglects fully the dissipation. Taking into account dissipation, the full trajectory and in particular the position of the center of the spiral described by $\mathcal{R}$ confirm $\theta = 90 ^\circ$ with the same accuracy of $\pm 11 ^\circ$ as in the low frequency regime.

\section*{Discussion}

We have directly observed the collective excitations of two coupled chiral edge channels at
filling factor 2 and demonstrated that it consisted in an antisymmetric neutral mode and a symmetric mode carrying the charge.
By creating selectively a charge density wave at
frequency $f$ in the outer edge and measuring the current
transferred to the inner one, we have observed oscillations as a
function of frequency that reflect the phase shift between the
charge and neutral modes. The minima of these oscillations
correspond to integer values of the ratio $l/\lambda_n$
between the propagation length and the wavelength of the neutral
mode. From these measurements, we have deduced the dissipation and
dispersion relation of the neutral mode, $k_{n}(\omega)$. We have observed two non-dispersive regimes, corresponding to a phase velocity $v_n(0)=4.6 \pm 0.6 \times 10^4$ m.s$^{-1}$ at low frequency and $v_n(\infty)=2.3 \pm 0.3 \times 10^4$ m.s$^{-1}$ at high frequency. Comparing our results with various models of inter-edge interactions, our
results show that edge channel propagation differs from the ideal
Luttinger limit of dissipationless propagation with short range
interaction, but rather agrees with a model of dissipative
channels coupled through a long range interaction. Dissipation
could be caused by the internal structure of compressible edges
leading to a coupling of EMP to acoustic modes
\cite{Aleiner94,Han97}  but a non ambiguous diagnosis will require
further investigation.

\section*{Methods}

\subsection*{Sample description}

\begin{figure}[h!]
\centering\includegraphics[width=1\textwidth]{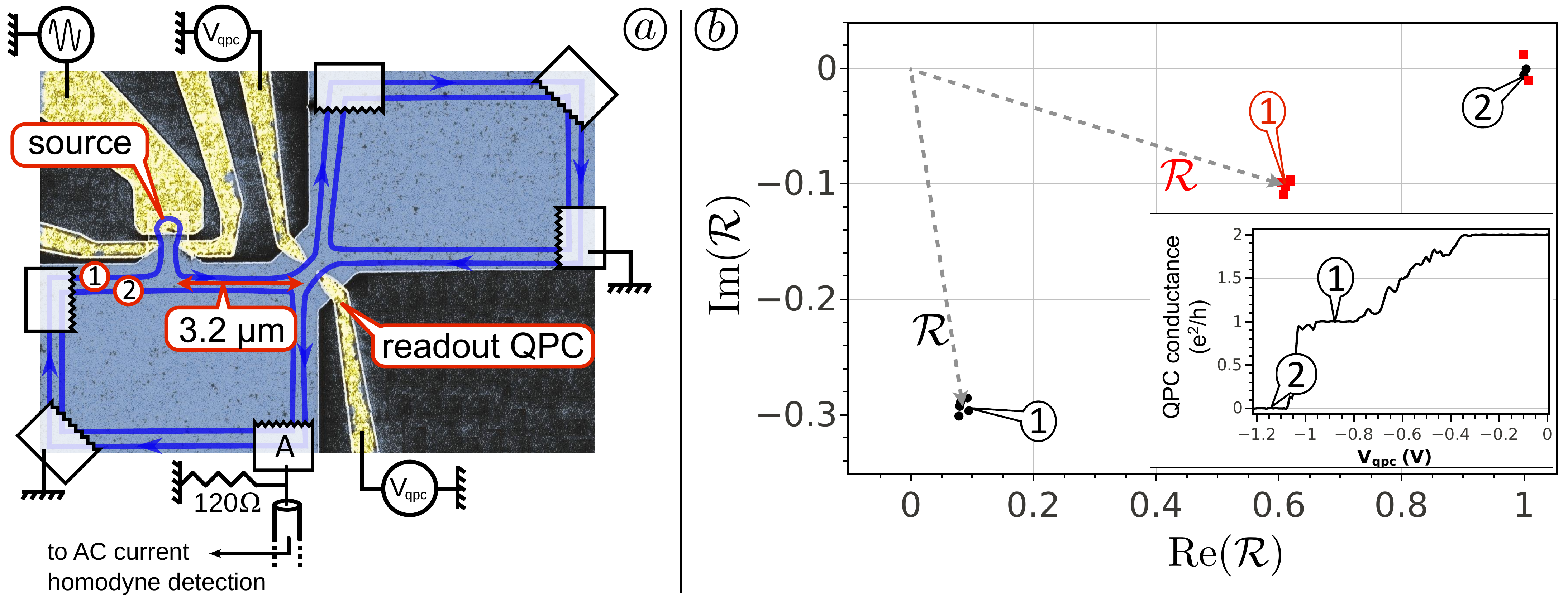}
\caption{ {\bf Schematics of the sample and principle of the experiment}\\(a) Schematic illustration of the
experiment based on the SEM picture of the sample. The two edge
states of filling factor $2$ are depicted in blue. A mesoscopic capacitor used as the source is
capacitively coupled to the outer channel only, and the EMP is generated by a sine drive of variable frequency
$f=\frac{\omega}{2\pi}$. The source is placed $3.2\pm0.4\, \mu$m before a
quantum point contact whose reflection can be varied so as to
enable selective readout of the current in both edge states. The
figure shows the setup in configuration 1 when channel 2 only is
reflected.  (b) Ratio $\mathcal{R}$ for
$f=1300$ MHz (black dots) and 5500 MHz (red
squares) in the complex plane. The principle of measurement is
illustrated : $\mathcal{R}$ is calculated from the ratio of the
total current measured in configuration 2 and the current in the
inner channel (configuration 1). The grey arrow is the complex
vector representing $\mathcal{R}$ in the complex plane. (Inset) DC
measurement of the conductance of the QPC, as a function of the
gate voltage $V_{qpc}$, illustrating configuration 1 and 2. } \label{fig1}
\end{figure}

\begin{figure}[h!]
\centering\includegraphics[width=0.9\textwidth]{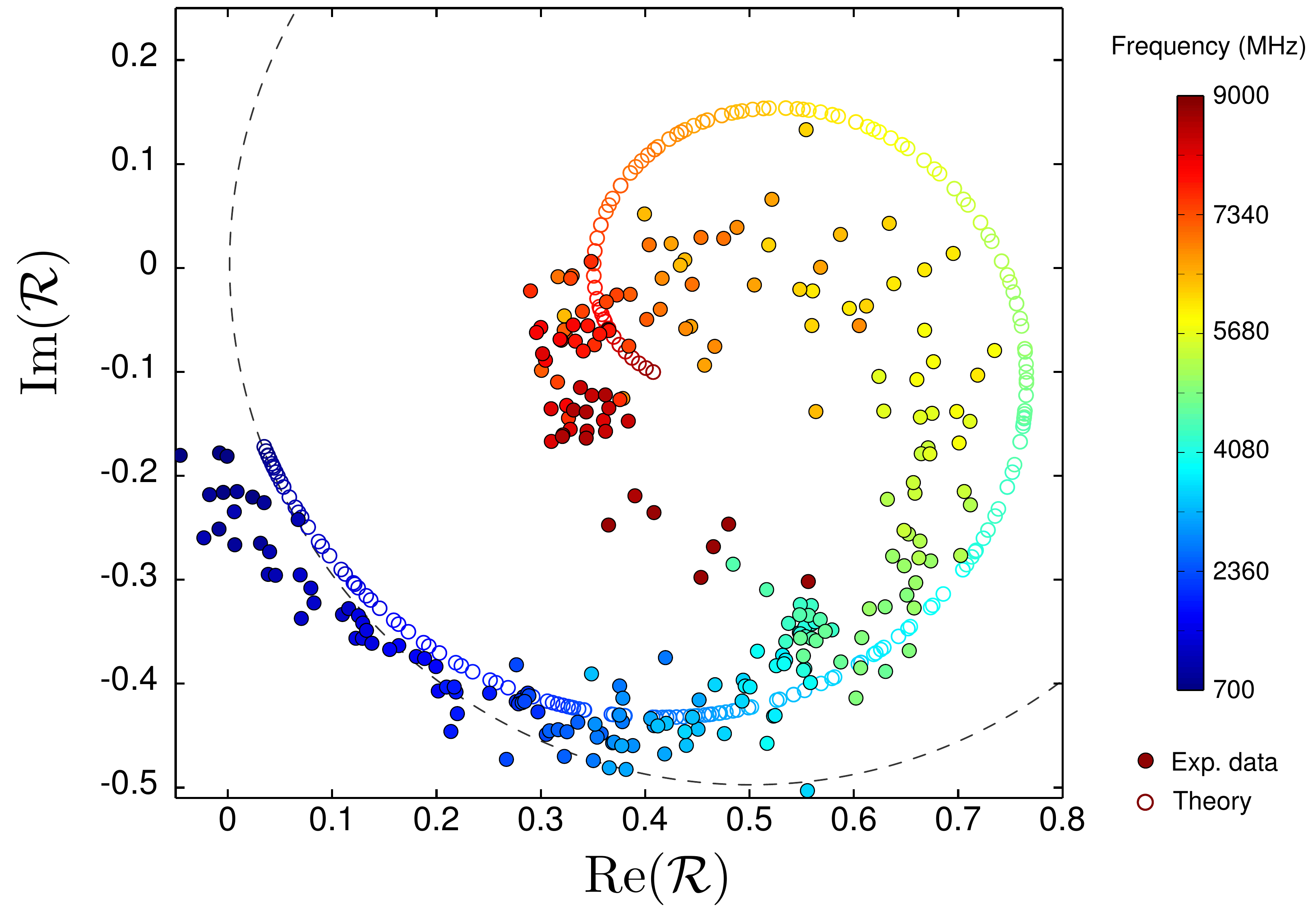}
\caption{{\bf Inter-channel oscillations}\\
Complex ratio $\mathcal{R}$ in the complex
plane. The color code indicates the frequency $f$ of the excited
EMP.  Experimental data (colored dots) are compared with
simulations. The grey dashes show a simulation of both short and
long range model, without relaxation. Colored hollow circles present a
simulation of the long range model with relaxation, for
parameters $\tilde\tau=35$ ps, $\tau_q=124$ ps (such that $\tau=
 81$ ps), and $\tau_r=4.1$ ps. }
\label{fig2}
\end{figure}

\begin{figure}[h!]
\centering\includegraphics[width=0.7\textwidth]{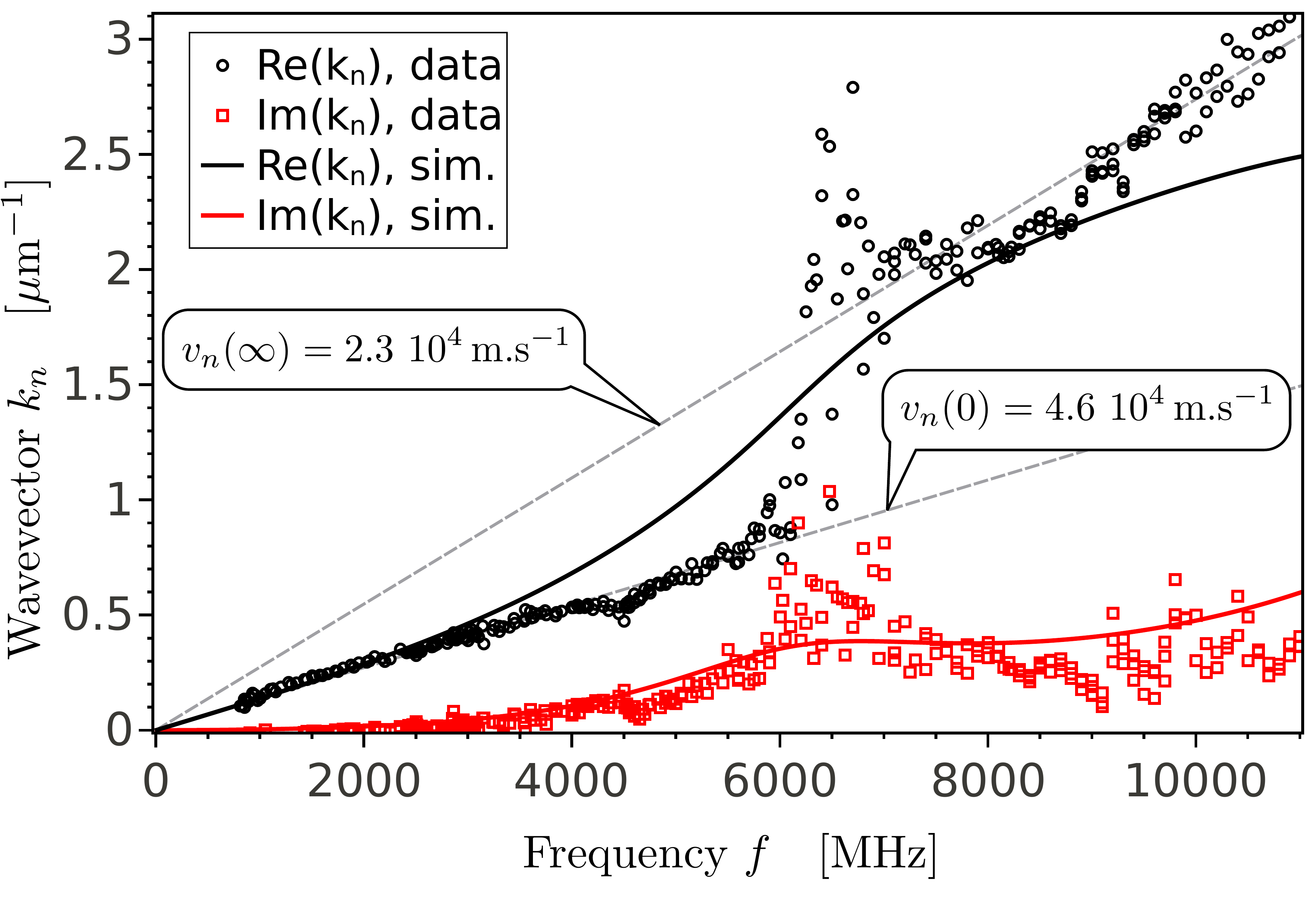}
\caption{{\bf Dissipation and dispersion of the neutral mode}\\
 Real and imaginary parts of the wave vector
$k_n(\omega)=\omega/v_n(\omega)$. ${\rm Re}(k_n)$
exhibits two non-dispersive regimes: at low frequency ($f<6$ GHz),
$v_n(0)=4.6\times 10^4$ m.s$^{-1}$, whereas at high frequency
($f>7$ GHz), $v_n(\infty)=2.3\times 10^4$ m.s$^{-1}$. ${\rm
Im}(k_n)\neq0$ indicates damping. Simulations (in black and red line) are proposed with parameters $\tilde\tau=35$ ps, $\tau_q=124$ ps ($\tau=
 81$ ps), and $\tau_r=4.1$ ps.} \label{fig3}
\end{figure}

\begin{figure}[h!]
\centering\includegraphics[width=0.7\textwidth]{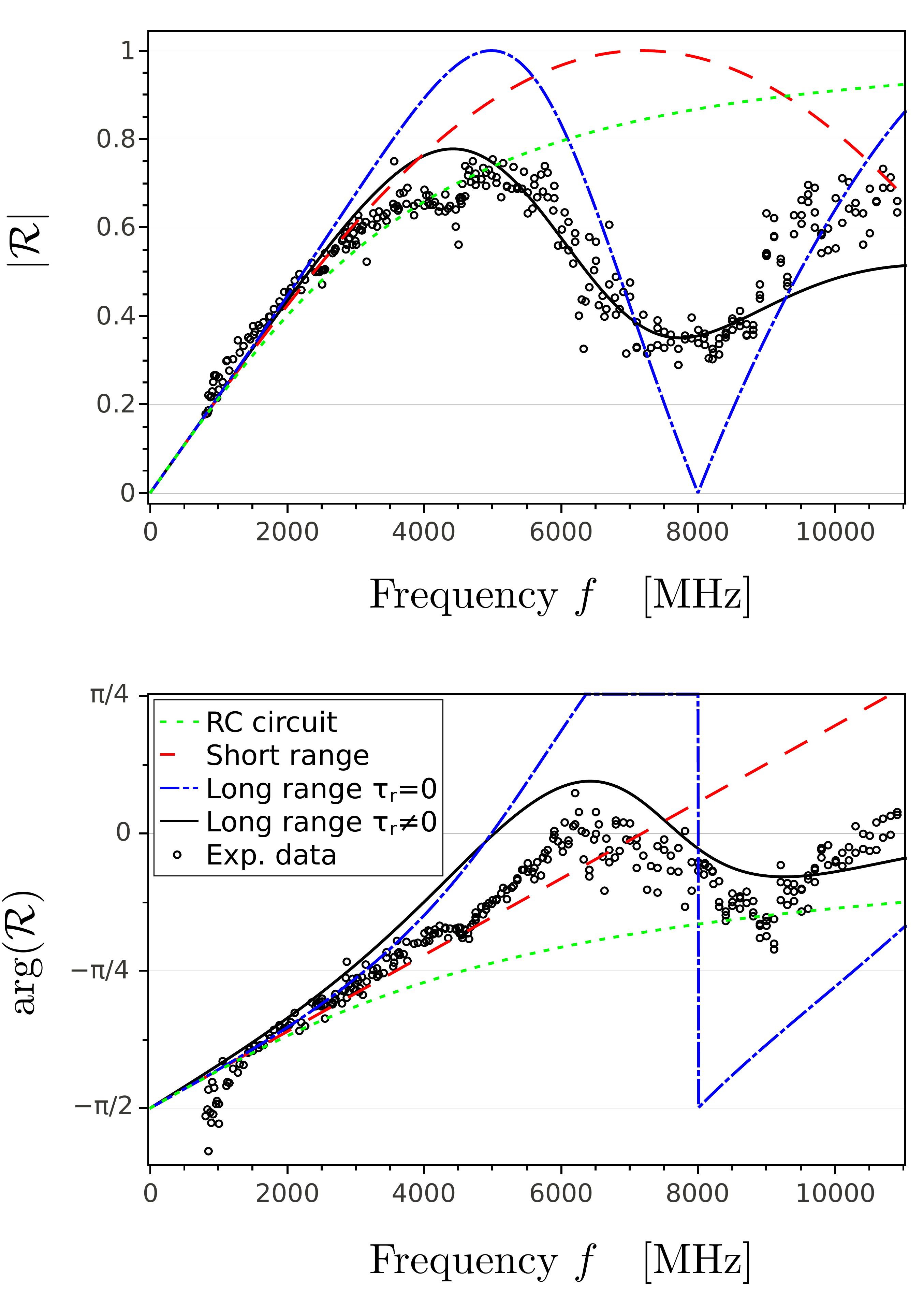}
\caption{{\bf Data-model comparison}\\
 $|\mathcal{R}|$ and  $\arg (\mathcal{R})$ as
a function of drive frequency $f$. Experimental data (black
circles) are compared with RC-circuit (green small dashes), short
range model (red dashes), long range model without damping (blue
dash-dotted line), long range model with relaxation (black plain
line). Parameters used are $\tilde\tau=35$ ps, $\tau_q=124$ ps ($\tau=
 81$ ps), and $\tau_r=4.1$ ps.} \label{fig3bis}
\end{figure}

\begin{figure}[h!]
\centering\includegraphics[width=0.8\textwidth]{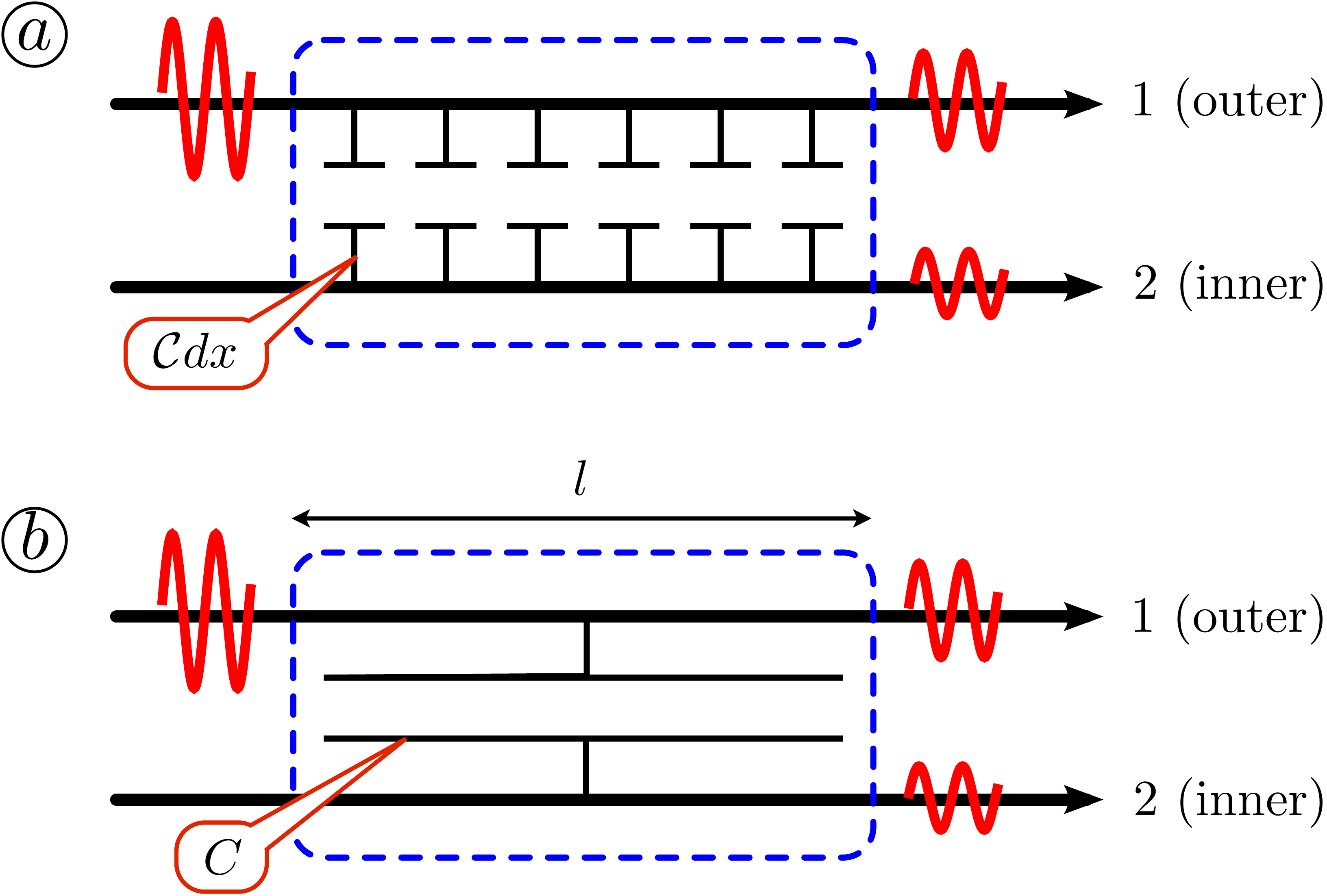}
\caption{{\bf Schematics of the zero range and long range interaction model. }\\
 For the zero range model (a), the charge densities $\rho_i(x,\omega)$
are locally coupled to the electrostatic potential $u_j(x)$ by the capacitance matrix per unit length,
$\rho_i(x,\omega)=\mathcal{C}_{ij}u_j(x,\omega)$. For the long range model (b), the potential in each edge $u_j$ is supposed to be uniform
and coupled to the total charges $q_i$ by the capacitance matrix $q_i=C_{ij}u_j$} \label{fig5}
\end{figure}

\begin{figure}[h!]
\centering\includegraphics[width=0.8\textwidth]{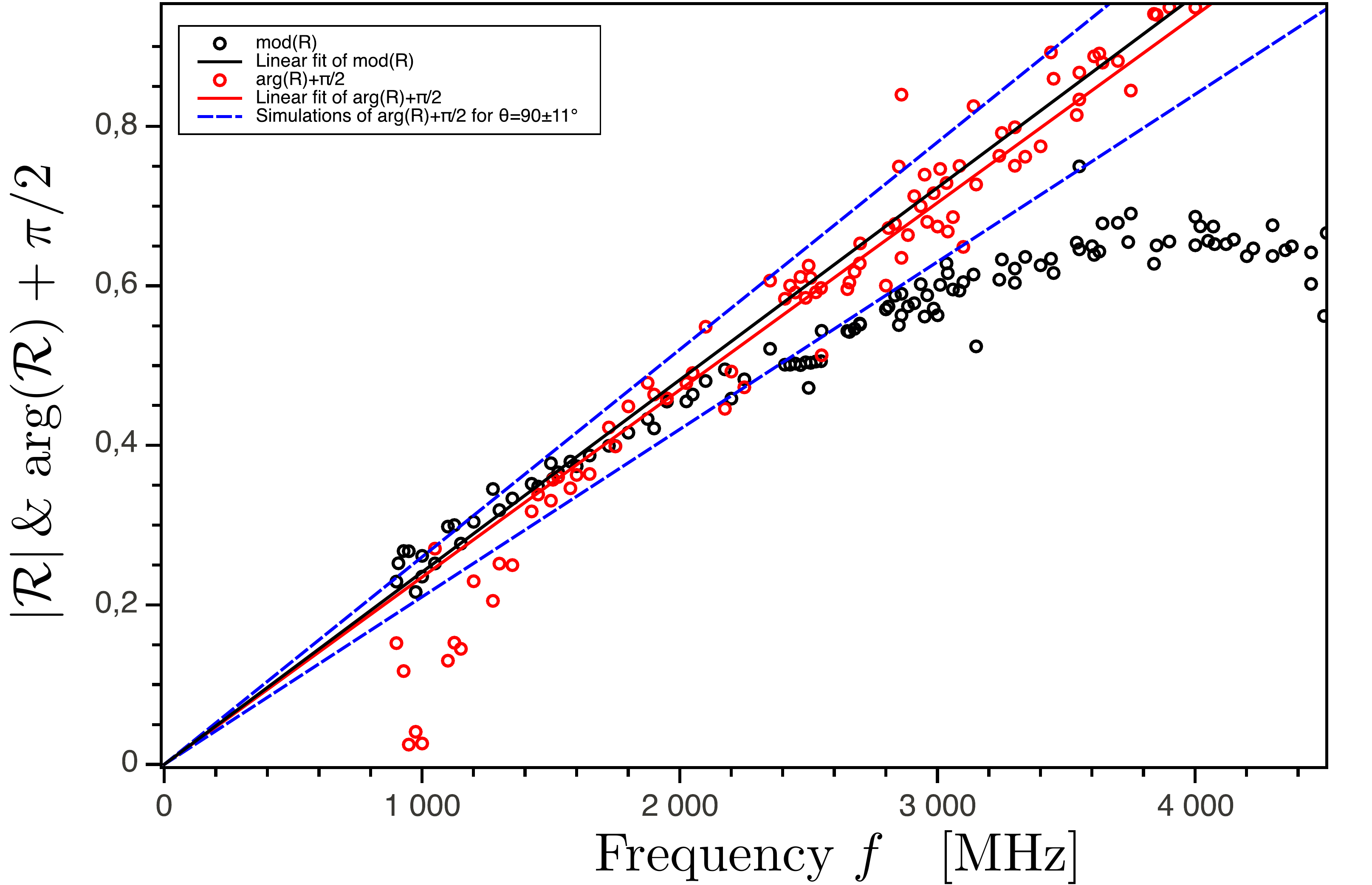}
\caption{{\bf Test of the nature of eigenmodes}\\
In the low-frequency regime, $|\mathcal{R}|$ and $\arg(\mathcal{R})  + \pi/2$ are presented respectively
in black and red circles, and fitted with linear functions (black and red lines).
The slope of $|\mathcal{R}|$ is obtained with a $10 \%$ accuracy, defining two bounds drawn in blue dashes associated to $\arg(\mathcal{R}) + \pi/2$ for $\theta=79^\circ$ and $\theta=101^\circ$. Data points fall between these bounds confirming $r=0 \pm 0.1$.
} \label{fig6}
\end{figure}

The sample is realized in a standard GaAs/Ga(Al)As two dimensional electron gas
located 100 nm below the surface, of density $n=1.8\, 10^{15}\ {\rm m^{-2}}$ and mobility $\mu={2.4\, 10^6\, \rm cm^2 V^{-1}s^{-1}}$.
The sample is then patterned using e-beam lithography and chemical etching of the heterojunction, and by deposition of
metallic gates at the surface. The electron gas is contacted using Gold/Germanium ohmic contacts schematically represented as white squares on Fig.\ref{fig1}(a).

The sample is placed in a strong magnetic field $B=3.65$
T so as to reach a filling factor $2$ in the bulk (see Fig.\ref{fig1} (a)). A driven mesoscopic capacitor (described in
references \cite{Buttiker1993a,Parmentier2012}) is used to selectively inject current in the outer edge channel (labeled 1).
The mesoscopic capacitor comprises a small portion of the electron gas (of submicronic size), called a quantum dot, capacitively coupled to a metallic top gate, see Fig.\ref{fig1} (a). A quantum point contact is used to fully transmit the outer edge channel (1) inside the dot
while the inner channel (2) is fully reflected. A sine drive of
frequency $f=\frac{\omega}{2\pi}$ is applied on the metallic top gate
deposited on top of the dot, so that an EMP of frequency $f$ is
capacitively induced in the outer channel, carrying a current
$i_1(0,\omega)$. The propagation takes place on a length $l\simeq
3.2\pm0.4\,\mu$m, during which channels are interacting. The EMP
then reaches a quantum point contact (QPC), that allows to reflect
or transmit selectively each edge channel. The reflected AC
current flows toward ohmic contact A, situated at distance
$L\simeq 60\, \mu$m from the QPC, and where the total current is measured. Note that the
propagation after the QPC, on length $L$, is irrelevant in the
analysis of the data: according to Eq.(\ref{Srho}), measuring
the total reflected current in ohmic contact A is equivalent, up
to a phase factor $e^{i\frac{\omega L}{v_\rho}}$, to measuring the
total current flowing right after the QPC.

The current is measured using a wideband room temperature homodyne
detection. A set of microwave filters and room-temperature low
noise amplifiers enables a proper measurement in the 0.7 to 11 GHz
range. Note that $\mathcal{R}$ does not depend on the total gain
of the amplifying detection scheme which varies considerably in
the studied frequency range. Each reported data point results from a simple averaging protocol,
and the statistical analysis is in good agreement with the observed dispersion of the experimental data.
At low frequency (below 1 GHz), the quality of our measurements is limited by the bandwidth of our
filters and amplifiers. At high frequency (above 10 GHz), it is limited for the same
reason, combined with increased attenuation of the output RF coaxial cables.

Experiments were performed in a dilution fridge of base temperature 50 mK. By performing Coulomb thermometry on the mesoscopic capacitor\cite{Gabelli2006a}, we have calibrated
the electronic temperature to $T_{el}=100\pm10$ mK.

\subsection*{Elements of theory}

In the integer quantum Hall regime, edge magnetoplasmons in channel 1 (outer) and 2 (inner) are described by a
chiral bosonic field $\phi_i(x,t)$ ($i=1,2$). In this approach, the two edge channels are bosonized in the spirit of Wen's description of quantum Hall edges as chiral Luttinger liquids \cite{Wen1990}. The current in channel $i$ is then
determined by $i_i(x,t)=\frac{e}{\sqrt{\pi}}\partial_t\phi_i(x,t)$, while the charge density
is $\rho_i(x,t)=-\frac{e}{\sqrt{\pi}}\partial_x\phi_i(x,t)$. Both are related via the current conservation equation.
In Fourier space, the motion of the chiral field $\phi_i(x,\omega)$ along the edge obeys:
\begin{eqnarray}
\big(-i\omega+\gamma(\omega)+v_i \partial_x\big)\phi_i(x,\omega)&=&\frac{e\sqrt{\pi}}{h}u_i(x,\omega) \label{eqphi}
\end{eqnarray}
where $v_i$ is the velocity of edge channel $i$ in the absence of interactions and $u_i$ the potential in edge channel $i$. The term $\gamma(\omega)$ models an intrinsic dissipation inside the edge states that accounts for the damping of the charge oscillations. Eq.(\ref{eqphi}) can be rewritten as a function of the current in edge channel $i$:
\begin{eqnarray}
\big(-i\omega+\gamma(\omega)+v_i \partial_x\big)i_i(x,\omega)&=&-\frac{i\omega }{R_K} u_i(x,\omega)
\end{eqnarray}

First, let us consider the non-damped case $\gamma(\omega)=0$. The short range description of the interaction can be obtained
by coupling locally the charge densities $\rho_i(x,\omega)$ to the local electrostatic potential $u_i(x,\omega)$ via
distributed capacitances $\mathcal{C}_{ij}$ (see Fig.\ref{fig5}): $\rho_i(x,\omega)=\mathcal{C}_{ij}u_j(x,\omega)$ where
$\mathcal{C}=-\mathcal{C}_{21}=-\mathcal{C}_{12}$ accounts for the coupling between channels whereas
$\mathcal{C}_{ii}$ describes intra-channel interactions in edge channel $i$.
These coupled equations are solved when working in the eigenbasis $\vec{u}_\pm$ that diagonalizes the velocity matrix,
$\mathcal{V}_{ij}=v_i \delta_{ij}+ \frac{e^2}{h}\mathcal{C}^{-1}_{ij}$. In the absence of inter-channel interaction, $\mathcal{C}^{-1}_{12}=0$, the channels are not coupled ($\theta=0, \vec{u}_+=(1,0),\vec{u}_-=(0,1)$)
but the velocities are renormalized by intra-channel interactions: $\mathcal{V}_{ii}=v_i + \frac{e^2}{h}\mathcal{C}^{-1}_{ii}$.
In the presence of inter-channel interactions ($\mathcal{C}^{-1}_{12}\neq 0$), new eigenmodes denoted by $+$ and $-$  and defined by Eqs.(\ref{EigenVecs}) to (\ref{EigenPhase}) with $\theta\neq0$ appear. The velocities $v_\pm$ and coupling angle $\theta$ are expressed as functions of the velocity matrix elements by:
\begin{eqnarray}
v_{\pm}&=&\frac{\mathcal{V}_{11}+\mathcal{V}_{22}}{2} \pm \sqrt{\frac{(\mathcal{V}_{11}-\mathcal{V}_{22})^2}{4}+ \mathcal{V}_{12}^2 }\\
\cos\theta&=&\frac{(\mathcal{V}_{11}-\mathcal{V}_{22})/2}{\sqrt{(\mathcal{V}_{11}-\mathcal{V}_{22})^2/4+\mathcal{V}_{12}^2}}\rm\ with\ \theta \in[0,\pi] \end{eqnarray}

As a consequence of the zero range of the interaction, the velocities $v_\pm$ are $\omega$-independent.
Note that the domain $\theta \in[0,\pi/2[$ corresponds the expected situation where, in the absence of inter channel interaction,  the outer edge channel velocity is greater than the inner one, $\mathcal{V}_{11} > \mathcal{V}_{22}$. The scattering matrix describing the coupled
propagation can then be straightforwardly calculated, yielding Eqs.(\ref{S21Theta}), (\ref{SsumTheta}), (\ref{RTheta}).
The charge ($+\to\rho$) and neutral ($-\to n$) eigenmodes are recovered for $\theta = \pi/2$, which always occurs for identical edge channels, $\mathcal{V}_{11}=\mathcal{V}_{22}$ but also for strong enough inter-channel interaction, $\mathcal{V}_{12} \gg \frac{\mathcal{V}_{11}-\mathcal{V}_{22}}{2}$.
As demonstrated above, this limit corresponds to the experimental situation. In this case, the velocity becomes $v_\pm= \overline{v}+ \frac{1}{R_K (\overline{\mathcal{C}} \pm \mathcal{C}) }$, with $\overline{v}=\frac{v_1+v_2}{2}$
and $\overline{\mathcal{C}}=\frac{\mathcal{C}_{11}+\mathcal{C}_{22}}{2}$. The case $v_{\rho} \gg v_{n}$ corresponds to total influence between edge channels, $\overline{\mathcal{C}} \approx \mathcal{C}$ and
such that $v_{n} \approx \overline{v}+ \frac{1}{2R_K \mathcal{C} }$. At low enough frequency, this short range model should describe correctly the coupling between channels. As discussed in the paper,
this corresponds to the RC circuit description, $S_{21}(\omega) \approx -i \omega \tilde{\tau} (1+ i\omega \tilde{\tau})$, $\tilde{\tau} = R_{K} \frac{C C_q}{2C  + C_q }$ where $C=l\mathcal{C}$ is the
total geometrical capacitance and $C_q= l e^2/(h v)$ the total quantum capacitance.

At higher frequencies, a way to account for long range interaction is to assume that the potentials $u_i$ are uniform in the whole edge channel $i$,
and related to the total charges in the 1D wires : $q_i=C_{ij}u_j$ with $C=C_{21}=C_{12}$ (see Fig.\ref{fig5}). In this description, the effective interaction range is given
by the co-propagating length $l$ itself. The same calculations can be performed to calculate $S_{21}$ and $\mathcal{R}$ in full generality (even for $\gamma(\omega)\neq0$ as detailed below).
From now on, we assume that $\theta=\pi/2$ as demonstrated in this article. Taking into account the damping in this model ($\gamma(\omega)\neq0$), we now obtain:
\begin{eqnarray}
S_{21}(\omega)&=&\frac{1-e^{i\omega\tau_q}e^{-\gamma(\omega) \tau_q}}{2+\frac{i}{(\omega+i\gamma(\omega))\tau}(1-e^{i\omega \tau_q}e^{-\gamma(\omega) \tau_q})}\\
\tau&=&R_K C,\  \tau_q=R_K C_q=\frac{l}{v}
\end{eqnarray}
For $\gamma(\omega)=0$, $S_{21}$ also draws a circle of radius 1/2
centered on (1/2,0) in the complex plane, but in contrast to the
short range case, the $\omega$-dependence of the phase deviates
from the linear law $\theta(\omega) \propto \omega $, which shows
that the velocity $v_{n}(\omega)$ becomes frequency
dependent. This frequency dependence is related to the two
timescales $\tau$ and $\tau_q$ introduced by the model. At low
frequencies one recovers the RC circuit description with
$\tilde\tau =\frac{\tau\tau_q}{2\tau+\tau_q}$. For $\gamma \neq 0$
and choosing $\gamma(\omega) = \omega^2 \tau_r$, as mentioned above, the
RC-circuit description is modified : a resistor of
$R_r=\frac{R_K}{2}\frac{\tau_r}{\tilde\tau}$ is added in series
with the charge relaxation resistor $R_K/2$.

\section*{References}

\bibliographystyle{unsrt}

\section*{Acknowledgements}
We gratefully acknowledge C. Glattli, C. Grenier, U. Gennser, F. Pierre, I.P. Levkivskyi, E.V. Sukhorukov, G. Haack, C. Flindt and M. B\"{u}ttiker for fruitful discussions, and T. Kontos for careful reading of the manuscript. This work is supported by the ANR grant '1shot',
ANR-2010-BLANC-0412.

\newpage

\end{document}